\documentclass{JHEP3}

\usepackage{amssymb}
\usepackage{amsmath}
\usepackage{graphicx}

\title{Soft SUSY breaking and family symmetry}
\author{M. R. Ramage$^a$ and G. G. Ross$^{a,b}$ \\
$^{a}$Department of Physics, Theoretical Physics, University of Oxford,\\
1 Keble Road, Oxford OX1 3NP, U.K.\\
$^{b}$ Theory Group, CERN, 1211 Geneva 23, Switzerland \\ E-mail: \email{ramage@thphys.ox.ac.uk}}
\abstract{A spontaneously broken non-Abelian $SU(3)$ family symmetry can generate a realistic form for quark, charged lepton and neutrino masses and mixing angles. It also gives a new solution to the SUSY flavour problem by ensuring near family degeneracy of the soft mass SUSY breaking terms. However the need to generate large third generation fermion masses means that the group must be strongly broken to $SU(2)$ giving significant corrections to the third family squark and slepton masses. We investigate the phenomenological implications of such breaking and show that it leads to new solutions capable of fitting all present experimental measurements and bounds as well as the dark matter abundance.}
\keywords{suy, ssm, sub, cos}
\preprint{hep-ph/0307389}

\begin{document}

\section{Introduction}

The origin of fermion masses and mixings is perhaps the most pressing of the
questions left unanswered by the Standard Model. Particularly noticeable is
the difference between the mixing angles in the quark and lepton sectors. In
the quark sector the mixing angles are small. However recent measurements of
neutrino oscillation have shown that the mixing of the atmospheric neutrinos
is consistent with being bi-maximal with equal components in the $\nu _{\mu
},$ $\nu _{\tau }$ directions while that of the solar neutrinos is
consistent with being tri-maximal with equal components in the $\nu _{e},$ $%
\nu _{\mu },$ and $\nu _{\tau }$ directions.

In \cite{King:2003rf} it was argued that this data suggests the existence of an
underlying non-Abelian family symmetry capable of relating the coupling of
the Higgs to different families. To demonstrate this an $SU(3)$ family
symmetry, the largest consistent with an underlying $SO(10)$ Grand Unified
symmetry, was constructed and shown to be capable of describing both the
quark and lepton masses and mixings. In order to generate the third
generation quark and charged lepton masses the family symmetry must be
strongly broken to $SU(2)$. The first two generation quark and charged
lepton masses are generated by a second stage of breaking which preserves a
discrete subgroup of the $SU(3)$ having matrix elements which are equal in
the $2$ and $3$ directions. With this breaking scheme, the effective Yukawa
couplings constrained by additional Abelian symmetries leads to mixing
angles in the up and down quark (and charged lepton) sectors which are small
and in agreement with the measured values. Although the Dirac mass of the
neutrinos has the same general form as that of the quarks and leptons the
light neutrinos have large mixing angles. This follows from the see-saw
mechanism with sequential domination of the right-handed neutrinos. In this
the near bi-maximal mixing comes from the correlation in the $2$ and $3$
directions of the second stage of breaking, a direct consequence of the
underlying $SU(3)$ symmetry. The near tri-maximal mixing of the solar
neutrino also follows from this vacuum alignment.

As stressed in \cite{King:2003rf} an important byproduct of such a family
symmetry is that, while unbroken, it guarantees the degeneracy of a family
of squarks or sleptons in a given representation of the Standard Model. As a
result it provides a new solution to the \textquotedblleft
family\textquotedblright\ problem, the need to have the squarks and charged
sleptons nearly degenerate to suppress flavour changing neutral currents and
to suppress CP violating effects in dipole electric moments. This solution
eliminates the need to appeal to the alternative solutions which communicate
supersymmetry breaking to the visible sector via various \textquotedblleft
mediator\textquotedblright\ mechanisms, including gravity mediation, gauge
mediation and anomaly mediation.

In this paper we will study the phenomenological implications of the $SU(3)$
family symmetry solution to the family problem. We concentrate on the
minimal case where an underlying Grand Unified symmetry guarantees the
degeneracy of the squarks and sleptons at the unification scale. However the
need to generate the third generation of quark and lepton masses requires
that there is strong breaking of this degeneracy for the third generation of
squarks and sleptons. Although we motivate this study in the context of a
specific implementation of a family symmetry model, it seems likely that the
structure applies more generally Any theory of fermion masses must
distinguish the third family from the light families and this is likely to
have an effect on the third sfamily masses too. We will show that the
splitting of the third family of sfermions following from the dominant
breaking of the family symmetry leads to significant change in the resulting
phenomenology. In particular we find a new class of solutions which satisfy
all the current bounds on supersymmetric states, have gauge coupling
unification and allow for radiative electroweak breaking, are consistent
with present measurements of $b\longrightarrow s\gamma $ and the anomalous
magnetic moment of the muon and have an LSP abundance which generates the
observed dark matter abundance.

The paper is organised as follows. In Section \ref{spectrum} we briefly
review the expectation for sfermion and Higgs masses in a supersymmetric
theory with an $SU(3)$ family symmetry. In Section \ref{method} we discuss
the various components of the global fit and the method used to perform the
renormalisation group flow and the analysis of the dark matter abundance.
The results are presented in Section \ref{results}. Finally in Section \ref%
{summary} we discuss the implications of the results and summarize.

\section{The sparticle spectrum\label{spectrum}}

In a theory with an underlying $SO(10)$ symmetry the soft supersymmetry
breaking masses of the states in a single family will be degenerate in the
absence of $SO(10)$ breaking. When the symmetry is extended to include a
non-Abelian $SU(3)$ family symmetry this degeneracy applies to soft
supersymmetry breaking masses of all the squarks and sleptons. If, in
addition, one assumes the Higgs scalars have the same initial mass one
obtains the Constrained Minimal Supersymmetric Standard Model (CMSSM)
spectrum often assumed in supergravity models, see for example, \cite{Inoue:1982pi,Hall:1983iz,Soni:1983rm,Nilles:1984ge,Haber:1985rc,Bartl:2001wc,Falck:1985aa,Bertolini:1990if}.

Of course, in the case of the non-Abelian family symmetry solution to the
family problem, there is no symmetry reason for assuming the degeneracy of
the Higgs scalars. The effect of breaking this degeneracy has been
extensively explored in \cite{Ellis:2003eg,Ellis:2002iu,Blazek:2002ta,Cerdeno:2004zj}. Here we wish to explore further
differences between the CMSSM and the non-Abelian symmetry solution to the
family problem. These arise when the GUT and family symmetries are broken
and are dependent on the pattern of symmetry breaking.

In \cite{King:2003rf} the full $SO(10)$ symmetric model was not constructed
but it was assumed the $SO(10)$ was broken close to the unification scale to 
$SU(4)\otimes SU(2)_{L}\otimes SU(2)_{R}.$ If this is the dominant breaking
effect the result is that the Pati Salam group $SU(4)$ preserves the
degeneracy of the up squarks and sneutrinos and of the down squarks and
sleptons in a given $SU(2)_{L}\otimes SU(2)_{R}.$ representation. Combined
with the $SU(3)$ family symmetry this means that there is degeneracy of the $%
SU(2)_{L}(SU(2)_{R})$ doublets of squarks and sleptons families as in the
CMSSM but there may be breaking between the left- and right- handed states.

Another possibility is that the dominant breaking is that of the $SU(3)$
family symmetry when giving the third family of fermions their masses. In
this case an underlying $SO(10)$ symmetry would guarantee the degeneracy of
all the states in a given family, but the breaking of $SU(3)$ to $SU(2)$
means that the degeneracy between the first two families and the third
family is lost. If $SO(10)$ is broken instead to $SU(5)$ the degeneracy
maintained is between the right-handed down squarks and the slepton doublets
and, separately, between the right-handed up squarks, the squark doublets
and the right-handed charged slepton. However, unlike the CMSSM, these two
groups of states can have different masses.

Of course both effects are likely to be present in a realistic theory but
the analysis of the general case is very difficult due to the large number
of parameters introduced. In this paper we shall explore the implications of
the second possibility where the dominant breaking is of the family symmetry
but the members of an $SO(10)$ multiplet remain degenerate at the
unification scale. We shall explore the more general possibilities elsewhere
but we choose to start with this simple case as the breaking of the family
symmetry breaks the solution to the family problem and, as discussed above,
can lead to significant changes in flavour changing and CP violating
processes.

In the model discussed in \cite{King:2003rf} the dominant soft supersymmetry
breaking mass terms come from the $D-$term $m_{0}^{2}\left( \psi _{i}\right)
^{\dagger }\psi _{i}|_{D}$ where $\psi _{i}$ are the quark and lepton
supermultiplets, triplets under the $SU(3)$ family symmetry, and $m_{0}^{2}$
is the supersymmetry breaking mass scale in the visible sector. This clearly
leads to degeneracy between the three families (we do not need to specify
the mediator sector as the degeneracy follows from the family symmetry). The
origin of the splitting between family multiplets comes from the D-terms $%
m_{0}^{2}\left( \psi _{i}\phi ^{i}\right) ^{\dagger }\psi _{j}\phi
^{j}/M^{2}|_{D}$ where the fields $\phi ^{i}$ are the ($SU(3)$ antitriplet)
fields which break the family symmetry. $M$ is the mass of the messenger
communicating family symmetry breaking to the quarks and leptons. The
dominant breaking comes from the field $\phi _{3}$ which has a vacuum
expectation value (vev) $\left\langle \phi _{3}^{3}\right\rangle =a.$ This
gives a relative mass difference of $O(a/M)$ to the third generation. In 
\cite{King:2003rf} this was chosen close to unity $a/M\simeq 0.4$, its
precise value ($<1)$ being undetermined. Breaking of $O(0.1)$ at the
unification scale between the third and first two generations must be
compared with the bounds on the splitting between families coming from
flavour changing neutral currents (FCNC). Such $O(1)$ breaking of the third
family gives a reduced effect on flavour changing bounds due to two effects.
Firstly radiative corrections from gauge interactions increase the average
quark and slepton masses at low scales without amplifying the breaking
between \ families, thus reducing the relative breaking effects. Secondly
the mixing of the third family to the light generations is small. As a
result the dominant breaking of the family symmetry is consistent with the
precision bounds coming from flavour changing processes \cite{Gabbiani:1996hi}.

The second breaking of the family symmetry is due to the field $\phi _{23}$
which has a vev $\left\langle \phi _{23}^{3}\right\rangle =\left\langle \phi
_{23}^{2}\right\rangle =b.$ The corresponding soft mass breaking term gives
a relative mass difference of $O(b/M)$ between the first two generations.
The fit to the light quark spectrum requires $b/M=O(0.02)$ in the down
sector and $O(0.003)$ in the up sector. Such splitting is within the bounds
of $O(1\%)$ on the breaking of the first two generations coming from FCNC.

Finally what about the remaining soft supersymmetry breaking parameters? The 
$A$ terms arise from $m_{0}W|_{A}$ ($W$ is the superpotential) \ which are
generated on supersymmetry breaking in the visible sector through a $%
m_{0}\theta \theta $ spurion ($\theta $ is the superspace coordinate). They
are the same in the effective theory coming from an underlying family
symmetry as in the CMSSM. This is because they arise from the effective Yukawa
couplings which are responsible for fermion masses and the choice of family
symmetry breaking has been dictated by the need to get viable fermion masses
as is assumed in the CMSSM.

We shall investigate the phenomenological implications of the breaking of
the family degeneracy in the soft SUSY breaking masses coming from the
pattern of symmetry breaking needed to generate the fermion masses. In
practice the breaking between the first two families is a negligible change
to the CMSSM boundary conditions but the splitting of the third family can
be significant. Although we have motivated this study in the context of a
specific model, the range of splitting we consider is dictated by the
observed fermion mass spectrum and so is likely to be the same for any
theory with an underlying family symmetry ordering the fermion masses which
is consistent with the FCNC bounds.

\section{Calculation and Constraints\label{method}}

In what follows we investigate the phenomenological implications of the
modification of the CMSSM spectrum discussed above which corresponds to the
case of a broken family symmetry. In particular we compare the case of a
degenerate scalar spectrum at the unification scale to one in which the mass
squared of the third generation of squarks and charged sleptons is allowed to vary
by up to 20\%. 

We use \texttt{SOFTSUSY} v.1.8.7~\cite{Allanach:2001kg}, one of several publicly available codes, to calculate the sparticle spectrum and mixings. The code has been augmented to include our family symmetry-inspired boundary conditions and a routine for the calculation of the SUSY contribution to the muon anomalous magnetic moment using the formulae in~\cite{Hisano:1996cp}. \texttt{SOFTSUSY} uses a bottom-up routine in which various low energy observables such as $M_Z$, fermion masses and gauge couplings are input as constraints in addition to the GUT scale boundary conditions. An iterative algorithm proceeds from an initial guess to find a set of sparticle masses and mixings consistent with the high and low scale constraints. We use full 2-loop renormalization group equations for the gauge and Yukawa couplings and the $\mu$ parameter. For the soft masses we use the full 1-loop RGEs and include the  2-loop contributions in the 3rd family approximation. Full details can be found in~\cite{Allanach:2001kg}. A comparison between \texttt{SOFTSUSY} and similar programs, for example~\cite{Porod:2003um,Djouadi:2002ze,Paige:2003mg} was made in~\cite{Allanach:2003jw} and one can directly compare the codes online at~\cite{Kramlweb:2004}. 

For the calculation of the neutralino relic density and $\mathcal{B}(b\rightarrow X_{s}\gamma )$ we use \texttt{micrOMEGAs} v.1.3.1~\cite{Belanger:2004yn}, linked to \texttt{SOFTSUSY} via an interface conforming with the Les Houches Accord~\cite{Skands:2003cj} standard that contains all the relevant parameters from \texttt{SOFTSUSY} necessary for the relic density calculation. For details of these calculations, see~\cite{Belanger:2004yn} and the papers on which they were based~\cite{Gondolo:1991dk,Edsjo:1997bg,Bertolini:1991if,Kagan:1998ym,Gambino:2001ew,Degrassi:2000qf,Chetyrkin:1997vx,Ciuchini:1998xe,Ciuchini:1998xy,Carena:1999py}. 

In our analysis we impose the following constraints:
\begin{itemize}
\item {\textbf{Direct searches}\newline
The following lower limits from LEP provide the strongest constraints on sparticle masses from direct searches~\cite{Eidelman:2004wy}: 
\begin{equation*} 
m_{\tilde{\chi}^{\pm}} \ge 103 \mathrm{GeV} \qquad m_{\tilde{e}_R} \ge 99 \mathrm{GeV}.
\end{equation*}
We include these lower bounds in our plots.}

\item {\textbf{Muon anomalous magnetic moment} \newline 
We include the $2\sigma$ bounds on the discrepancy between experiment and Standard Model theory assuming the latest results of the calculation based on $e^+ e^-$ data for the hadronic contribution~\cite{Akhmetshin:2003zn} and the most recent data from the BNL E821 experiment incorporating the results from negative muons~\cite{Bennett:2004pv}. We use the values from~\cite{Hagiwara:2003da} which include the recently recalculated $\alpha^4$ QED correction~\cite{Kinoshita:2004wi} and the most recent hadronic light-by-light contribution~\cite{Melnikov:2003xd}. Similar values were obtained by an independent calculation~\cite{Davier:2003pw}. However this second paper does not take the new theoretical results~\cite{Kinoshita:2004wi,Melnikov:2003xd} into account. From~\cite{Hagiwara:2003da}, 
\begin{equation*}
a_{\mu}^{exp} - a_{\mu}^{SM} = (24.5\pm9.0)\times 10^{-10},
\end{equation*}
where $a_{\mu} \equiv \frac{(g-2)_{\mu}}{2}$. We use the $2\sigma$ bound,
\begin{equation*}
6.5\times 10^{-10} < \delta a_{\mu} < 42.5\times 10^{-10},
\end{equation*}
as the allowed range of the SUSY contribution. Due to the inconsistency between these results and those obtained by using $\tau$ decay data, and taking into account the susceptibility to change of the measurement of the $e^+ e^-$ cross section~\cite{Akhmetshin:2003zn}, the $(g-2)_{\mu}$ constraint should perhaps be viewed more provisionally than the others. This is unfortunate since it is one of the most important, being the only one that unambiguously determines the sign of $\mu$.}

\item {\textbf{Branching Ratio} $\mathbf{\mathcal{B}(b\rightarrow X_{s}\gamma )}$ \newline
The most recent world average for the branching ratio is~\cite{Jessop:2002ha}
\begin{equation*}
\mathcal{B}(b\rightarrow X_{s}\gamma )_{exp}=(3.34\pm 0.38)\times 10^{-4},
\end{equation*}
while the current Standard Model theory value is~\cite{Bieri:2003jm}\footnote{This takes into account only those results that include the improved ratio $m_c^{\overline{MS}}(m_b/2)/m_b^{pole}$ as opposed to $m_c^{pole}/m_b^{pole}$ in the $\langle X_s \gamma |(\bar{s}c)_{V-A}(\bar{c}b)_{V-A}|b\rangle$ matrix element. For details see~\cite{Gambino:2001ew}.} 
\begin{equation*}
\mathcal{B}(b\rightarrow X_{s}\gamma )_{SM}=(3.70\pm 0.30)\times 10^{-4}.
\end{equation*}
We will use this Standard Model estimate of the theoretical error in our calculation as representative of the error to be expected in our calculation which includes both Standard Model and SUSY contributions. We do this by combining the experimental and theoretical errors in quadrature to obtain the following upper and lower bounds on the branching ratio at $2\sigma$:
\begin{equation*}
2.40\times 10^{-4}<\mathcal{B}(b\rightarrow X_{s}\gamma )<4.28\times 10^{-4}.
\end{equation*}
}

\item {\textbf{Neutralino dark matter} \newline
The analysis of the data from WMAP gives a best fit value for the matter density of the universe of $\Omega_{m}h^2 = 0.135^{+0.008}_{-0.009}$ and for the baryon density, $\Omega_{b}h^2 = 0.0224\pm 0.0009$~\cite{Spergel:2003cb}. This implies that the CDM density is
\begin{equation*}
\Omega_{CDM}h^2 = 0.1126^{+0.0161}_{-0.0181}
\end{equation*}
at the $2\sigma$ level. This can be an extremely stringent bound on the MSSM parameter space, especially in the case of small $\tan\beta$. However, for large $\tan\beta$ it is less restrictive due to the presence of the $A^0$ Higgs resonance, and much less so if we allow for a source of cold dark matter other than neutralinos such as axions, or some relic density enhancement mechanism such as non-thermal production of neutralinos (see~\cite{Profumo:2004at} and references therein for more examples). In these instances, the lower bound on $\Omega_{m}h^2$ can be neglected. We plot values for which 
\begin{equation*}
0.0945 < \Omega_{CDM}h^2 < 0.1287,
\end{equation*}
and indicate the allowed regions if we choose to discard the lower bound. We also plot the locus of points for which $m_{A^0} = 2m_{\tilde{\chi}_1^0}$ marking the position of the $A^0$ resonance.}

\item {\textbf{Lightest Higgs Mass} $\mathbf{m_{h^0}}$ \newline
We also display the contour 
\begin{equation*}
m_{h^0} = 114.1 \mathrm{GeV},
\end{equation*}
corresponding to the LEP bound on the lightest SM Higgs boson~\cite{Eidelman:2004wy} in the regions of parameter space where the lightest MSSM Higgs boson is Standard Model like, i.e. \mbox{$\sin(\beta - \alpha)$} is almost exactly equal to 1, where $\alpha$ is the mixing angle relating the mass eigenstates to the gauge eigenstates in the CP-even neutral Higgs sector. This condition applies throughout the parameter space we analyse here.} 

\item {\textbf{Correct EWSB / Tachyons / Higgs potential unbound from below}\newline
The boundary on which $|\mu|^2$ vanishes, marking the border of correct radiative electroweak symmetry breaking has been plotted. In the region where $|\mu|^2 < 0$ a global minimum of the two loop effective Higgs potential cannot be found. Similarly, any regions in which $m_{A^0}^2 < 0$, also signalling that the electroweak symmetry has not been broken correctly, have been excluded. Regions with tachyonic sfermions are likewise omitted.}  
\end{itemize}

\section{Results and Discussion\label{results}}

We first present our results for universal boundary conditions, i.e.
all scalar masses are set to $m_{0}$ and all gaugino masses to $m_{1/2}$ at
the GUT scale. Fig.~\ref{fig1} shows the $(m_{1/2},m_{0})$ plane of the
CMSSM for $\tan \beta =10$, $30$ and $50$ for plots (a), (b) and (c)
respectively. We set $\mu >0$ in accordance with the expectation for $(g-2)_{\mu }$, and $A_{0}=0$ for simplicity. Our plots can be seen to be in reasonable agreement with those in recent papers~\cite{Ellis:2003cw,Djouadi:2001yk,Baer:2003jb,Lahanas:2003yz,Roszkowski:2001sb,Barger:2001yy,Arnowitt:2002he}\footnote{Any discrepancies between the results shown in Fig.~\ref{fig1} and those of other papers are likely to be due to the differing approximations used to compute the sparticle spectrum and mixings and the neutralino relic density. One can find comparisons between various commonly used codes in ref.~\cite{Allanach:2003jw}.}

\FIGURE[tbp]{\includegraphics[width=0.551\textwidth]{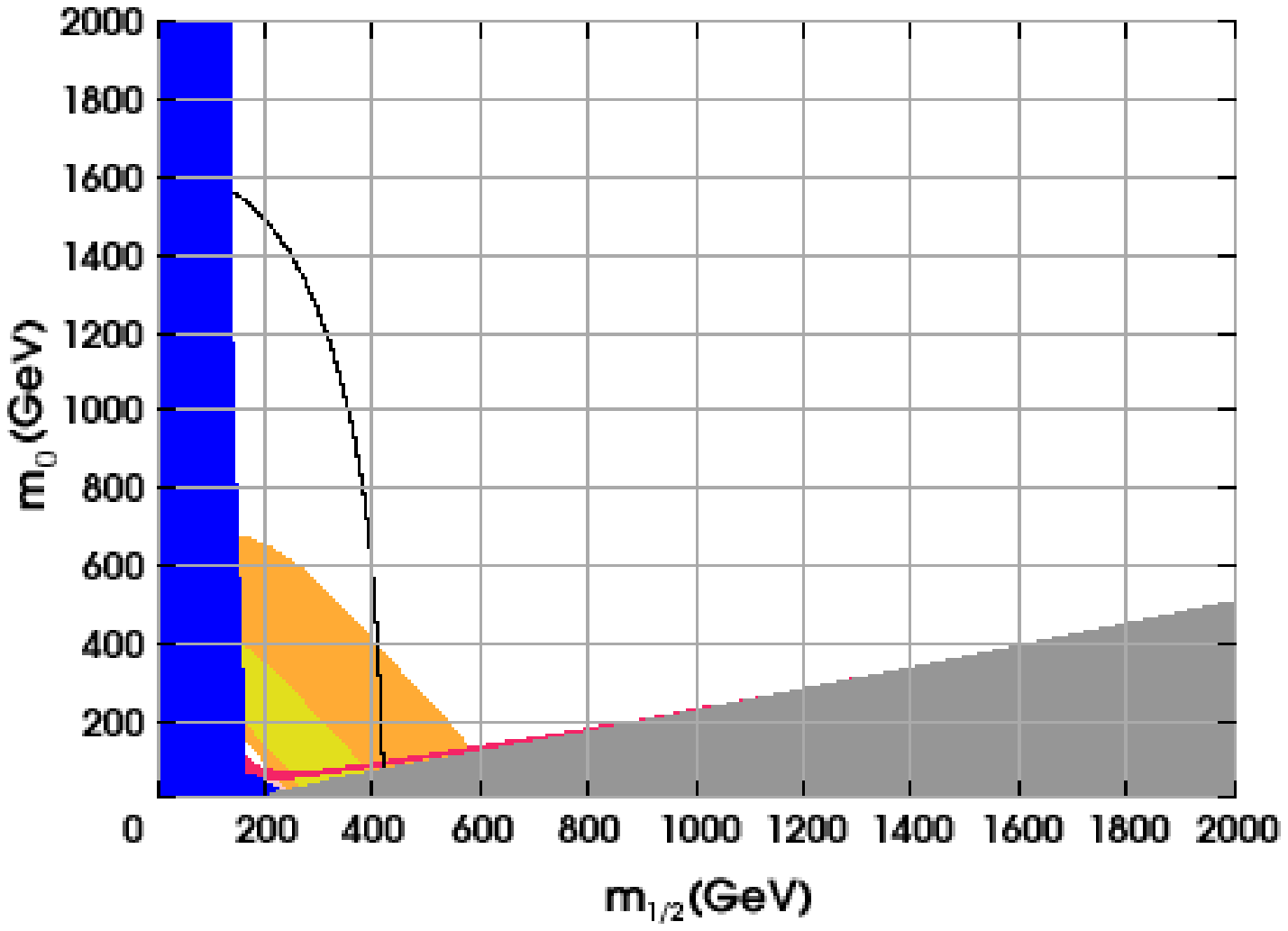} (a) 
\includegraphics[width=0.551\textwidth]{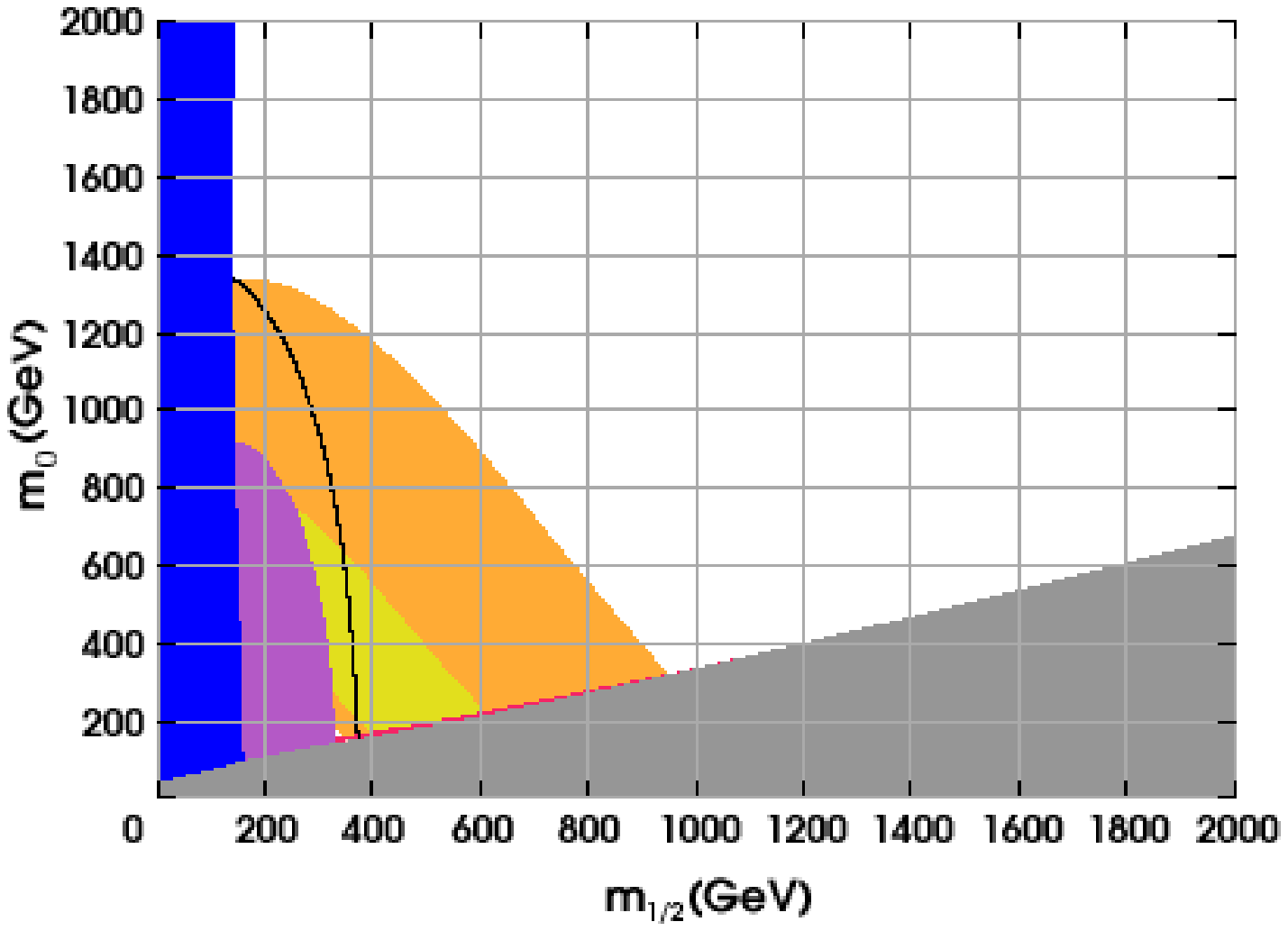} (b) 
\includegraphics[width=0.551\textwidth]{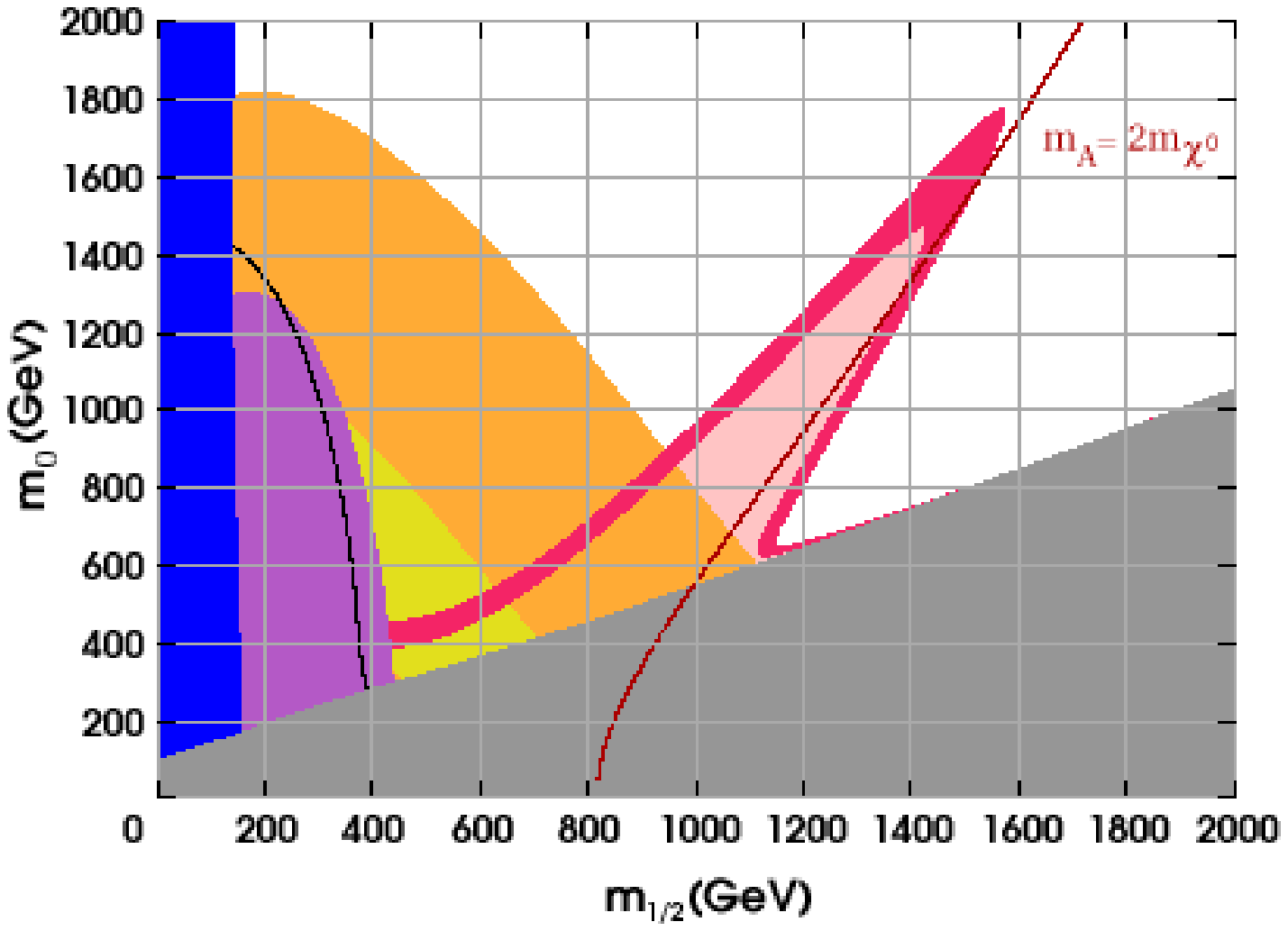} (c)
\caption{{\it The ${(m_{1/2},m_{0})}$  plane with ${\mu > 0}$, $A_0 = 0$ in the CMSSM for {\rm(a)} $\tan\beta = 10$, {\rm (b)} $\tan\beta = 30$ and {\rm (c)} $\tan\beta = 50$. The medium grey region is excluded because the LSP is a stau, the blue(v. dark grey) region is excluded by the LEP limits on sparticle masses and $\mathcal{B}(b \rightarrow X_s \gamma)$ is too small in the lilac(darkish grey) region; the black line corresponds to the contour $m_{h^0} = 114.1$ GeV; the red(dark grey) strip shows where $0.0945 < \Omega_{CDM} h^2 < 0.1287$ and the yellow(v. light grey) and orange(light grey) regions represent the $1\sigma$ and $2\sigma$ bounds on the region favoured by $(g-2)_\mu$. The position of the $A^0$ resonance where $2m_{\tilde{\chi}_1^0} = m_{A^0}$ is shown as a dark red (dark grey) line and regions satisfying only the upper bound on $\Omega_{CDM}h^2$ are shaded light pink (v. light grey).}}\label{fig1}}

We focus our discussion on the neutralino relic density. There is a narrow
band for each value of $\tan\beta$ where the bounds on $\Omega_{CDM} h^2$
and $(g-2)_\mu$ are satisfied. In this region, for $\tan\beta = 10$, $30$, the main annihilation channels for the neutralinos are t-channel
sfermion exchanges to leptons and quarks with coannihilations with
staus becoming important close to where $m_{\tilde{\chi}^0_1} = m_{\tilde{\tau}}$. For $\tan\beta =
50$ the main channels in the favoured region are again t-channel sfermion
exchanges, but also s-channel $A^0$ Higgs exchanges. Rapid annihilation via the $A^0$ dominates as the resonance is approached. Near to the band of parameter
space excluded by the chargino mass, for $\tan\beta = 30$, $50$, there is a
narrow filament of acceptable relic density in the favoured region
corresponding to the $h^0$ resonance. The focus point region~\cite{Feng:1999zg,Feng:1999mn,Feng:2000gh,Feng:2000bp} where $|\mu|$ becomes very small,
resulting in a large Higgsino component of the LSP, occurs at a higher value
of $m_0$ than shown in these plots. The boundary where $|\mu| = 0$ begins at $m_0 \simeq 2300$ GeV for $m_{1/2} = 
100$ GeV and $\tan\beta = 50$. In this sector of parameter space
annihilation to gauge bosons is enhanced and chargino coannihilation also
becomes important resulting in an acceptable relic density. However, this
region is well outside the range favoured by $(g-2)_{\mu}$.

We now compare these results with those predicted by the $SU(3)$ family
symmetry. Since we do not know the sign of the correction to the third family sfermion masses we consider two additional cases. Fig.~\ref{fig2} is the same plot as Fig.~\ref{fig1}, but with the
soft supersymmetry breaking sfermion masses squared taking the following form
at the GUT scale: 
\begin{equation*}
m^{2}_{\tilde{Q}}(m_G) = m^{2}_{\tilde{u}_R}(m_G) = m^{2}_{\tilde{d}_R}(m_G) = m^{2}_{\tilde{L}}(m_G) = m^{2}_{\tilde{e}_R}(m_G)= m^{2}_0\left(%
\begin{array}{ccc}
1 & 0 & 0 \\ 
0 & 1 & 0 \\ 
0 & 0 & 1-\delta m^2%
\end{array}%
\right),
\end{equation*}

and the same for Fig.~\ref{fig3}, but with 
\begin{equation*}
m^{2}_{\tilde{Q}}(m_G) = m^{2}_{\tilde{u}_R}(m_G) = m^{2}_{\tilde{d}_R}(m_G) = m^{2}_{\tilde{L}}(m_G) = m^{2}_{\tilde{e}_R}(m_G) = m^{2}_0\left(%
\begin{array}{ccc}
1 & 0 & 0 \\ 
0 & 1 & 0 \\ 
0 & 0 & 1+\delta m^2%
\end{array}%
\right)
\end{equation*}
where $\delta m^2 = 0.2$ is the correction coming from the $SU(3)$ family
symmetry. Although the model also predicts small corrections to the (1,1)
and (2,2) elements of the sfermion mass matrices, they have a negligible
effect on the phenomenology so we will ignore them.

\FIGURE[tbp]{\includegraphics[width=0.551\textwidth]{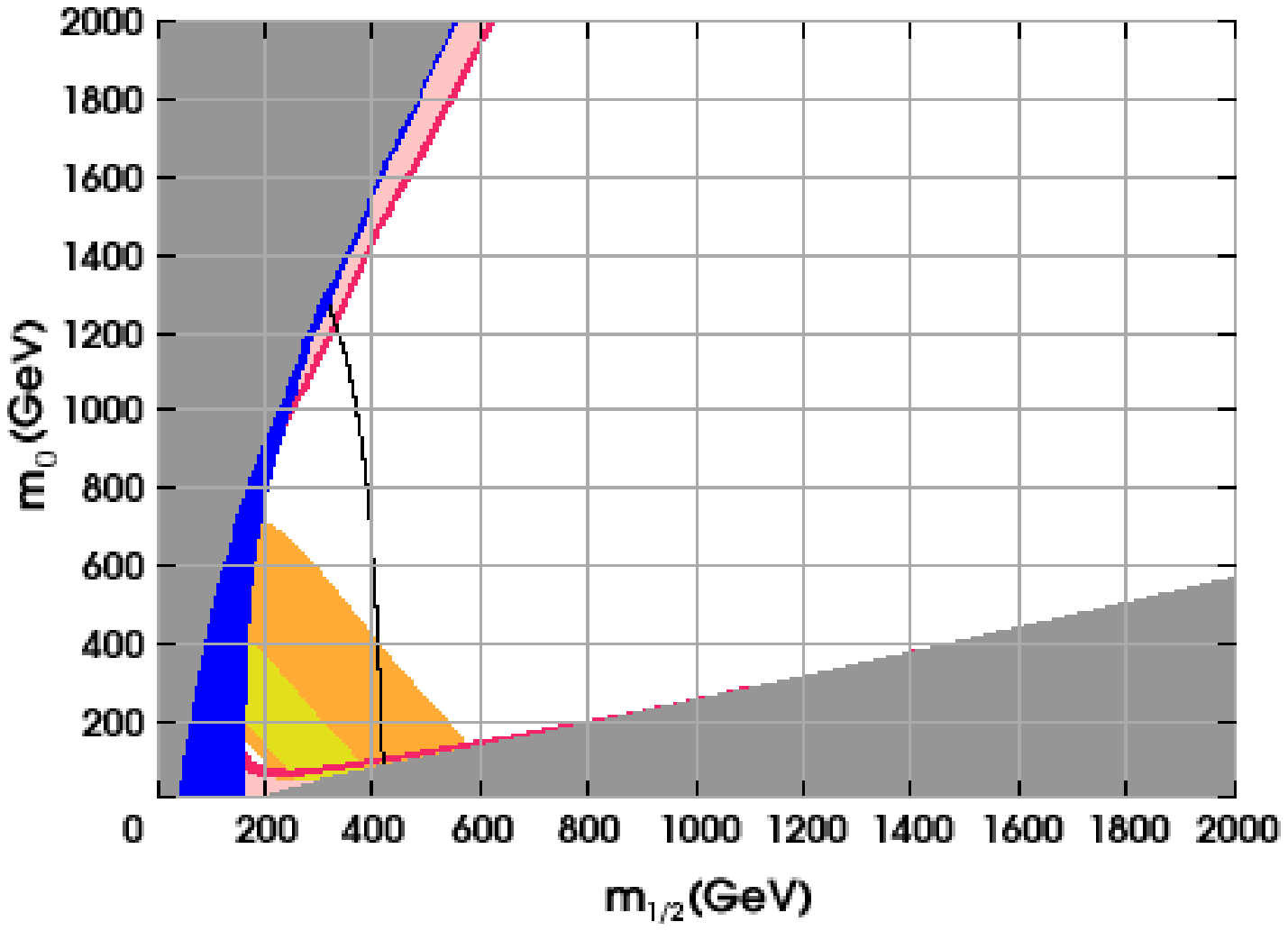} (a)
\includegraphics[width=0.551\textwidth]{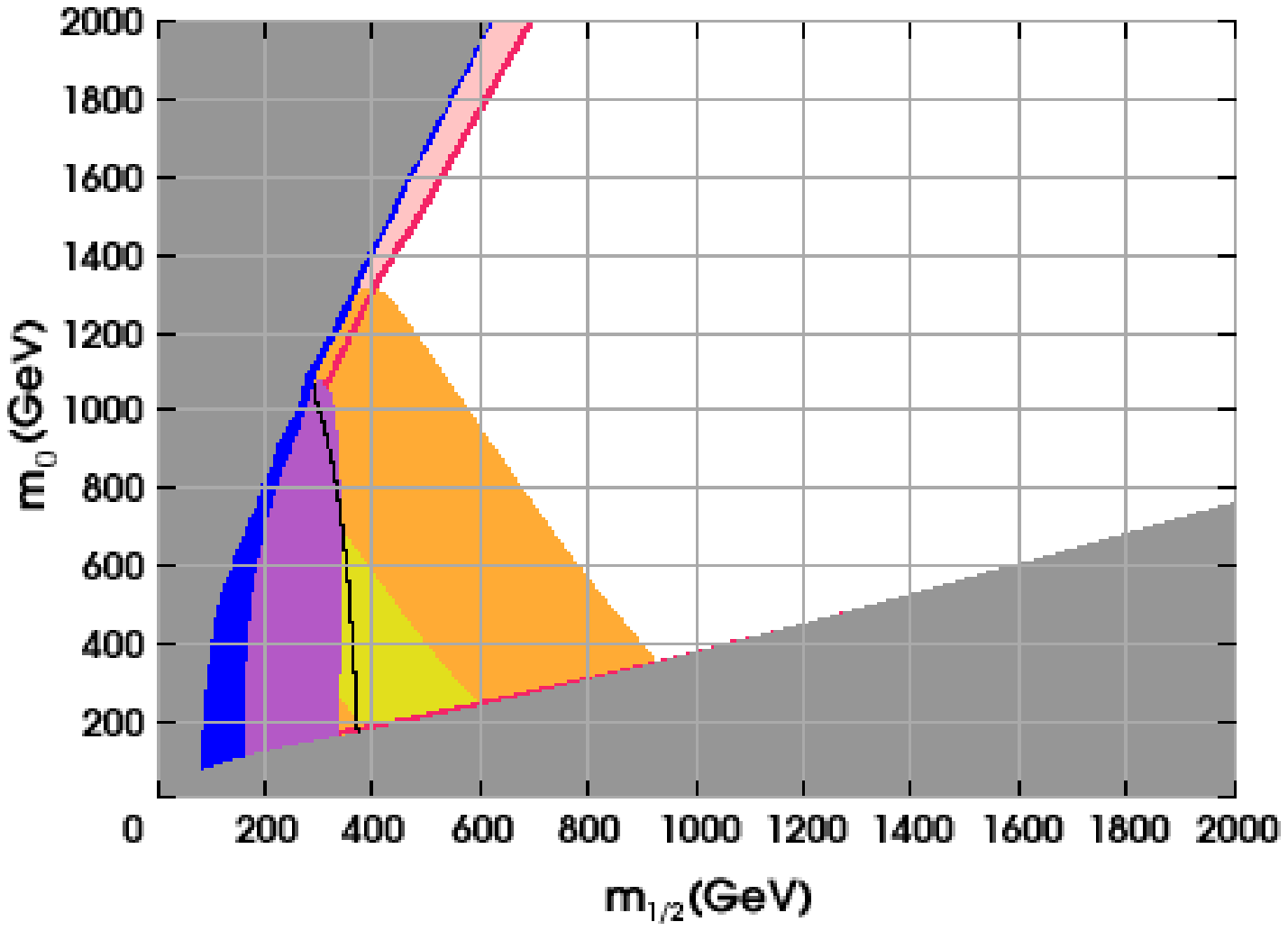} (b) 
\includegraphics[width=0.551\textwidth]{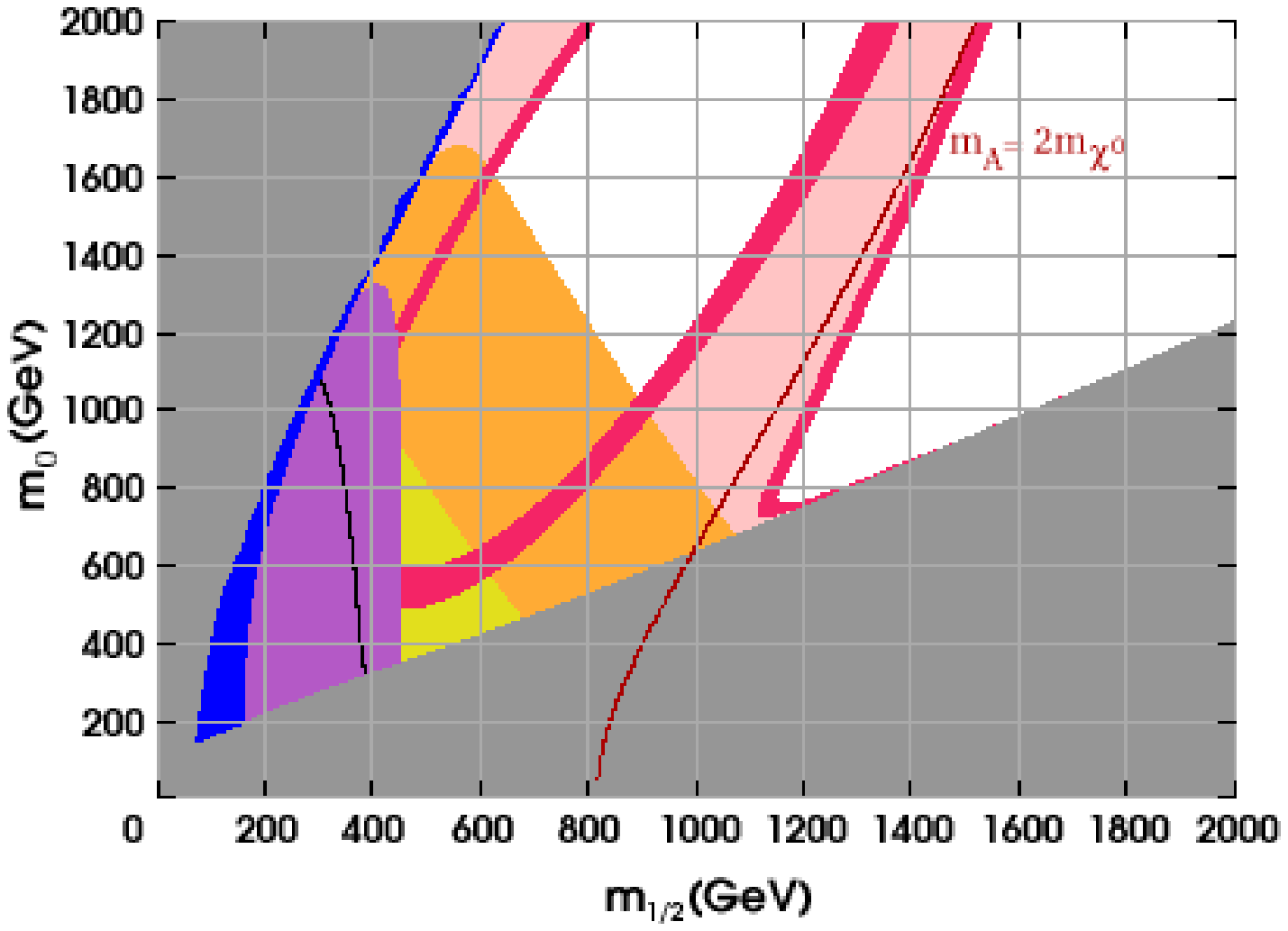} (c)
\caption{\textit{Same as Fig.~\protect\ref{fig1}, but with $m^{2}_{\tilde{Q}%
}(m_G) = m^{2}_{\tilde{u}_R}(m_G) = m^{2}_{\tilde{d}_R}(m_G)  = m^{2}_{\tilde{L}}(m_G) = m^{2}_{\tilde{e}_R}(m_G) = m_0^2 \mathrm{diag}(1,1,1-\protect\delta m^2)$. The new medium grey region on the left hand side is excluded by EWSB requirements, i.e. $|\protect\mu|^2 < 0$.}}\label{fig2}}

\FIGURE[tbp]{\includegraphics[width=0.551\textwidth]{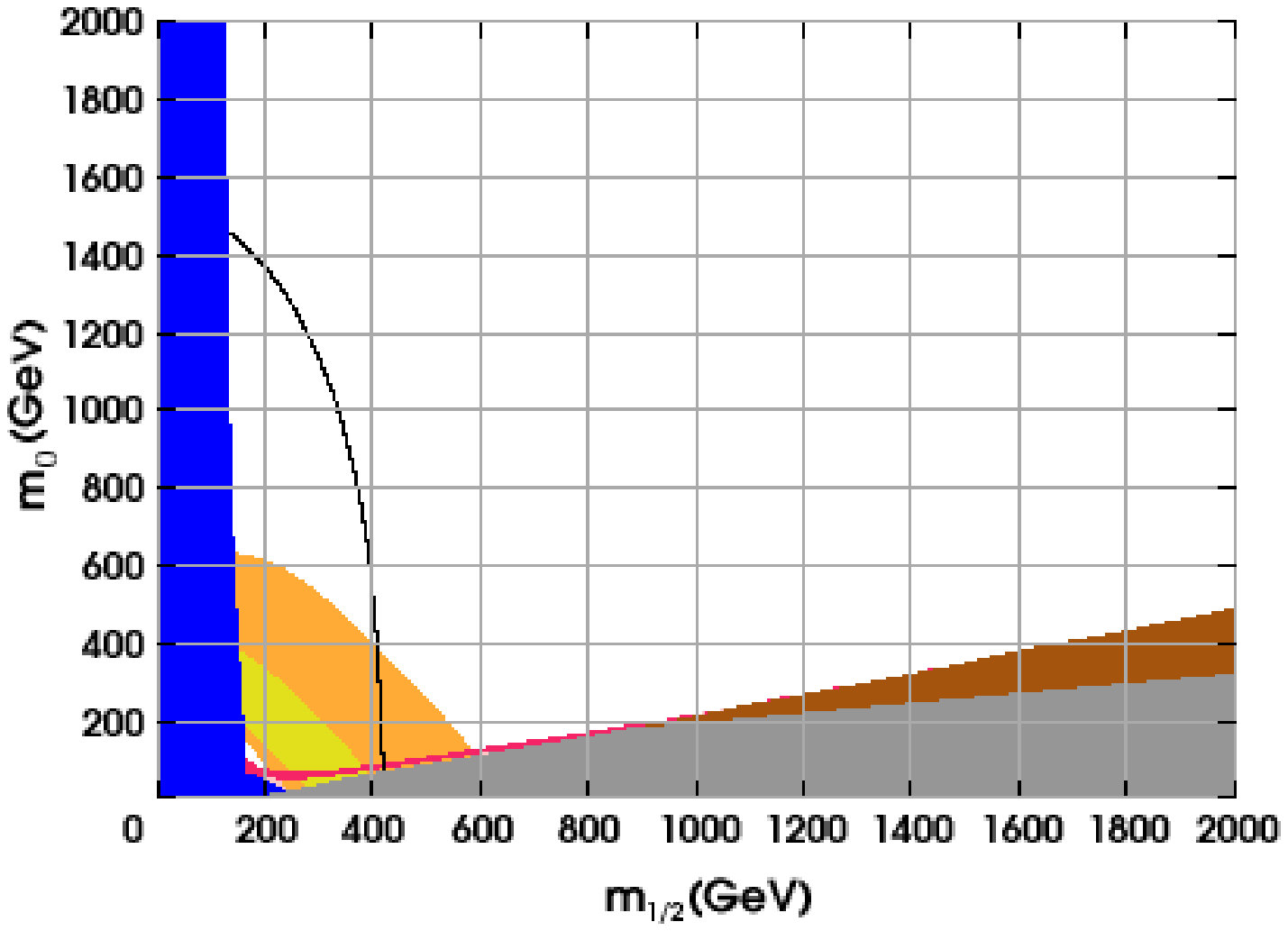} (a) 
\includegraphics[width=0.551\textwidth]{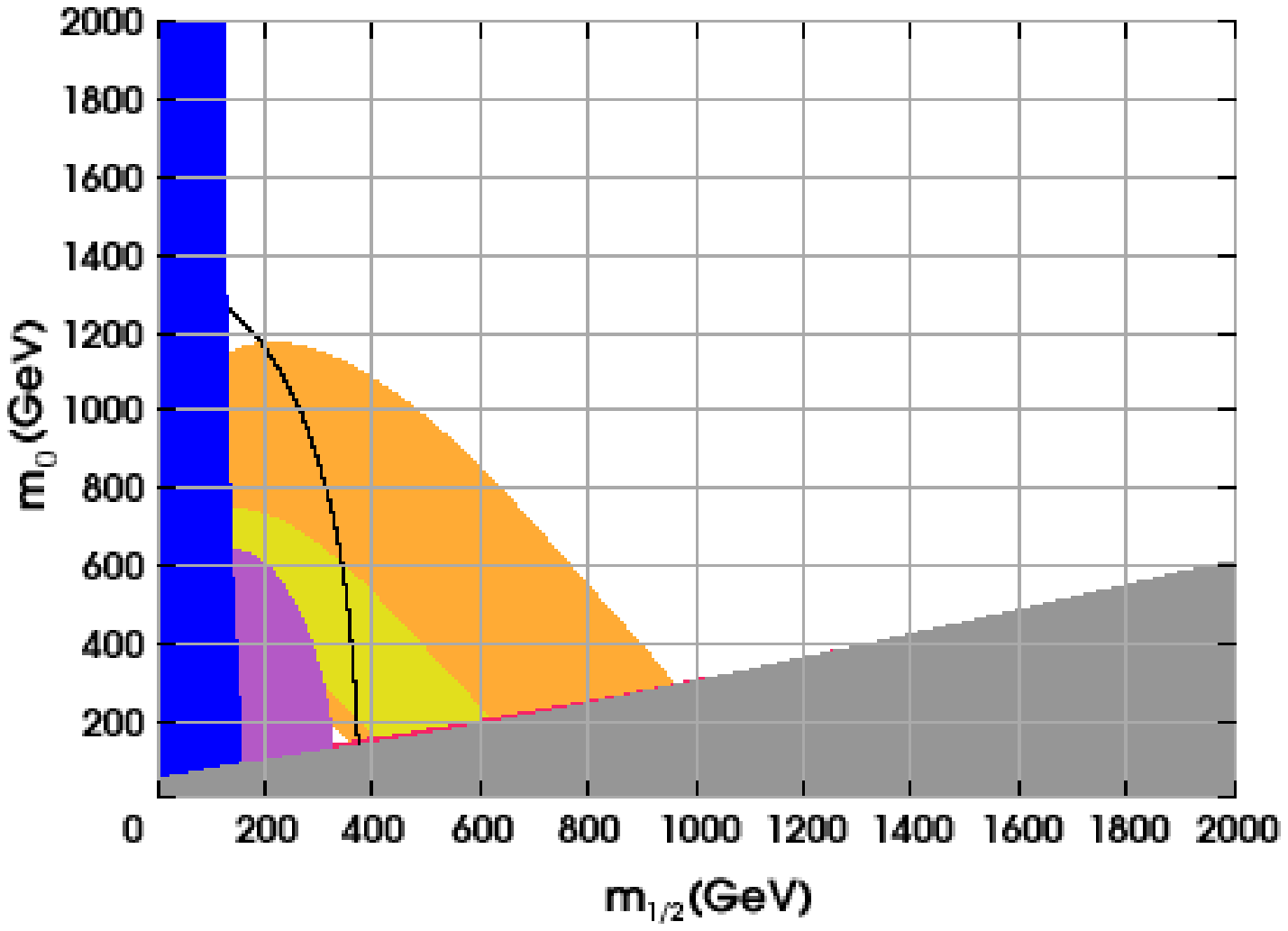} (b) 
\includegraphics[width=0.551\textwidth]{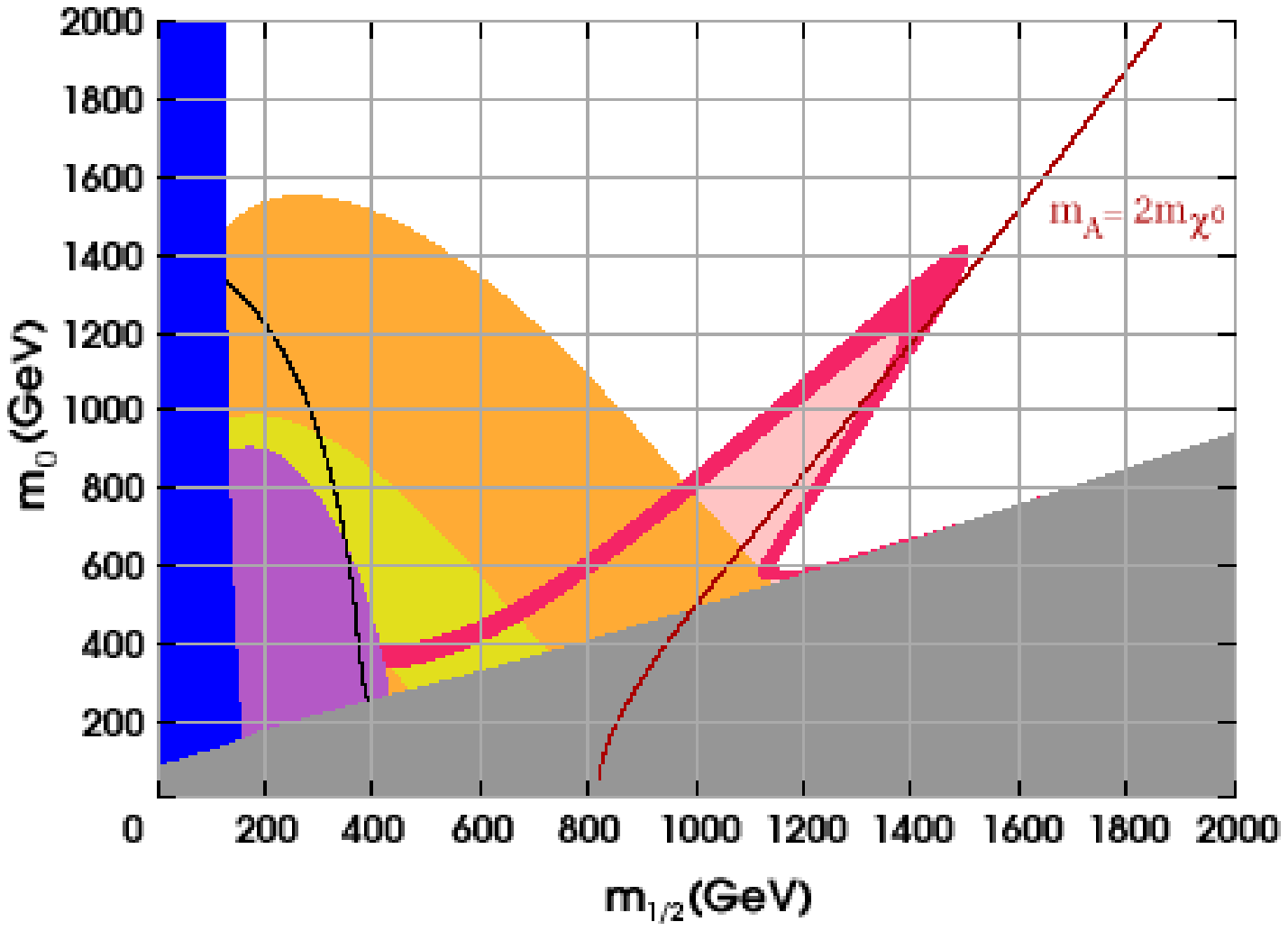} (c)
\caption{\textit{Same as Fig.~\protect\ref{fig1}, but with $m^{2}_{\tilde{Q}%
}(m_G) = m^{2}_{\tilde{u}_R}(m_G) = m^{2}_{\tilde{d}_R}(m_G)  = m^{2}_{\tilde{L}}(m_G) = m^{2}_{\tilde{e}_R}(m_G) = m_0^2 \mathrm{diag}(1,1,1-\protect\delta m^2)$. In the new brown(dark grey) region in {\rm (a)} the LSP is a smuon, not a stau. }}\label{fig3}}

From Figs.~\ref{fig2} and \ref{fig3}, it is clear that altering the universal
boundary conditions in this way has a relatively small effect on the regions
favoured by $(g-2)_\mu$ or on those excluded by $\mathcal{B}(b \rightarrow
X_s \gamma)$ or direct search constraints and we will not discuss these effects
further. Due to the variation of the slepton soft mass matrices there is a 
small change in the slope of the boundary $m_{\tilde{\chi}^0_1} = m_{\tilde{\tau}}$ in the three different cases. This has a rather small effect on the preferred region of parameter space. However, there is a very significant change in the region of
allowed electroweak symmetry breaking and the neutralino relic density in
the case of decreased third family soft sfermion masses as shown in Fig.~\ref{fig2}, especially for large $\tan\beta$. As we shall see, this is 
predominantly due to the change in the third family squark soft masses, and is relatively insensitive to changes in the slepton sector.

One of the main features of the plots in Fig.~\ref{fig2} is the large region
where the electroweak symmetry is not broken correctly. Here, $|\mu|^2 < 0$
is found when the minimisation conditions are applied to the scalar
potential indicating that an acceptable minimum cannot be found. The
boundary of this region corresponds to $|\mu| = 0$. The effect of decreasing
the third family sfermion soft masses on EWSB can be understood by
considering the expression for $|\mu|^2$ from the EWSB conditions and the
RGE for $m_{H_2}^2$. Including quantum corrections~\cite{Pierce:1997zz}, 
\begin{eqnarray}
|\mu|^2 &=& \frac{\overline{m}_{H_1}^2 - \overline{m}_{H_2}^2 \tan^2\beta}{%
\tan^2\beta - 1} - \frac{1}{2} m_{Z}^2 - \frac{1}{2} \Re\mathfrak{e}
\Pi_{ZZ}^T  \notag \\
&\simeq& \frac{\overline{m}_{H_1}^2}{\tan^2\beta} - \overline{m}_{H_2}^2 - 
\frac{1}{2}m_{Z}^2 - \frac{1}{2} \Re\mathfrak{e}\Pi_{ZZ}^T  \notag
\end{eqnarray}
since $\tan\beta \gg 1$. Here, $\overline{m}_{H_{i}}^{2}=m_{H_{i}}^{2}-t_{i}/v_{i}$
where $t_{i}/v_{i}$ are the tadpole contributions, and $\Pi _{ZZ}^{T}$ is the
transverse part of the $Z$ self-energy. This implies the following condition on $\overline{m%
}_{H_2}^2$: 
\begin{displaymath} 
-\overline{m}_{H_2}^2 \ge \frac{1}{2}m_{Z}^2 + \frac{1}{2} \Re\mathfrak{e}%
\Pi_{ZZ}^T - \frac{\overline{m}_{H_1}^2}{\tan^2\beta}, 
\end{displaymath}
which must be satisfied in order to obtain $|\mu|^2 \ge 0$.

Since we have $\tan ^{2}\beta \geq
100$, the $\overline{m}_{H_{1}}^{2}$ term is suppressed and therefore we
require the $\overline{m}_{H_{2}}^{2}$ term to be very small or negative for
successful EWSB. This can be achieved by large radiative corrections~\cite%
{Ibanez:1982fr}. The one-loop RGE for $m_{H_{2}}^{2}$ in the third family approximation is~\cite{Martin:1997ns} 
\begin{displaymath}
16\pi ^{2}\frac{dm_{H_{2}}^{2}}{d\log Q}=6|y_{t}|^{2}\left(m_{H_{2}}^{2}+(m_{%
\tilde{Q}}^{2})_{33}+(m_{\tilde{u}%
_{R}}^{2})_{33}\right)+6|a_{t}|^{2}-6g_{2}^{2}|M_{2}|^{2}-\frac{6}{5}%
g_{1}^{2}|M_{1}|^{2}.
\end{displaymath}%
One can see from this that by decreasing the squark mass squared parameters $(m_{\tilde{Q}}^{2})_{33}$ and $(m_{\tilde{u}_{R}}^{2})_{33}$,  
$m_{H_{2}}^{2}$ will be driven to relatively higher values at low scales. As
a result the range of $m_{0}$ for which the electroweak symmetry will be
broken successfully is reduced for a given $m_{1/2}$. A correlated consequence is that
the effects associated with the focus point region in the standard CMSSM are
pushed to much lower values of $m_{0}$ and a new strip of acceptable relic
density consistent with the constraints appears for $%
\tan \beta \gtrsim 30$ due to the increased Higgsino component of the LSP.

It should be noted that in this region the primary mechanism for neutralino
annihilation is {\it not} annihilation to massive gauge bosons or 
coannihilation with charginos, but annihilation to $b\bar{b}$ and 
$\tau\bar{\tau}$ via an s-channel $A^0$. The region where annihilation to gauge 
bosons or chargino coannihilation dominates is closer to the boundary where 
$|\mu|=0$ and here the relic density is too small to account for 
$\Omega _{CDM}h^{2}$. As one moves away from this boundary in the direction of 
increasing $m_{1/2}$ towards the region of favoured relic density two 
important things happen.
Firstly, the Higgsino component of the neutralino decreases significantly. 
As a result, since the amplitude for annihilation to gauge bosons is proportional to the 
square of their coupling to the Higgsino component of the neutralino, the 
cross-section rapidly drops. This is not the case for the annihilation amplitude featuring an s-channel $A^0$ boson because it only depends linearly on the coupling 
to the Higgsino component. Secondly, the mass difference between the lightest chargino and the LSP neutralino increases and coannihilation quickly becomes negligible. Although not at the $A^0$ 
resonance, relative to the case of no sfermion splitting, the amplitude for 
annihilation via the $A^0$ is enhanced for two reasons:

(i){Due to the larger Higgsino component of the neutralinos, their coupling 
to the $A^0$ is increased.}

(ii) {The mass difference $2m_{\tilde{\chi}^{0}_1}-m_{A^0}$ is far smaller than for the
usual focus point region at large $m_{0}$, thus enhancing the propagator.}

As $m_{1/2}$ is increased further, past the first band of acceptable relic density, there is a rise in $\Omega_{CDM}h^2$ caused by the decrease in Higgsino component of the LSP. However, in the case of $\tan\beta = 50$, $\Omega_{CDM}h^2$ quickly drops again as the $A^0$ resonance effects become important. This explains the existence of two bands of acceptable neutralino relic density in Fig.~\ref{fig2}(c) below the $A^0$ resonance for $m_0 \gtrsim 1200$ GeV.

For $\tan\beta \gtrsim 50$ another effect comes into play. As well as
decreasing with increasing $\tan\beta$, $m_{A^0}$ decreases with decreasing third family sfermion soft masses. Therefore the zone of parameter
space in which the $A^0$ resonance occurs is pushed to lower values of $m_{1/2}$.
 Moreover, this effect contributes to reason (ii) given in the
previous paragraph. When $\tan\beta \gtrsim 50$, $\Omega_{CDM}h^2 \le 0.0945$
for most of the region favoured by $(g-2)_\mu$ not excluded by other 
constraints. The reason for this reduction in $%
m_{A^0}$ can be understood from the formula for $m_{A^0}^2$ from the EWSB
conditions and the renormalisation group equations for $m_{H_1}^2$ and $%
m_{H_2}^2$.

From the EWSB requirements~\cite{Martin:1997ns}, 
\begin{eqnarray}
m_{A^0}^2 &=& \frac{1}{\cos 2\beta}\left(\overline{m}_{H_2}^2 - \overline{m}%
_{H_1}^2\right) - m_{Z}^2 - \Re\mathfrak{e} \Pi^T_{ZZ}(m_{Z}^2)  \notag \\
&& - \Re\mathfrak{e} \Pi_{AA}(m_{A^0}^2) + \frac{t_1}{v_1}\sin^2\beta + \frac{%
t_2}{v_2}\cos^2\beta \nonumber
\end{eqnarray}
where $\Pi_{AA}$ is the $A^0$ self energy. The dominant term is the one
containing the soft Higgs masses so $m_{A^0}^2 \sim m_{H_1}^2 - m_{H_2}^2$.
Comparing the relevant terms in the renormalisation group equations for
these parameters~\cite{Martin:1997ns} 
\begin{displaymath}
16\pi^2 \frac{dm_{H_2}^2}{d\log Q} = 6|y_t|^2\left(m_{H_2}^2 + (m_{\tilde{Q}%
}^2)_{33} + (m_{\tilde{u}_{R}}^2)_{33}\right) + \ldots
\end{displaymath}
\begin{displaymath}
16\pi^2 \frac{dm_{H_1}^2}{d\log Q} = 6|y_b|^2\left(m_{H_1}^2 + (m_{\tilde{Q}}^2)_{33} + (m_{\tilde{d}_{R}}^2)_{33}\right) + 2|y_{\tau}|^2 \left(m_{H_1}^2 + (m_{\tilde{L}}^2)_{33} + (m_{\tilde{e}_{R_{}}}^2)_{33} \right) + \ldots
\end{displaymath}
one can see that, due to the large top Yukawa coupling, any change in the
soft sfermion masses will have a larger effect on the renormalisation group
equation for $m_{H_2}^2$ than for $m_{H_1}^2$. Indeed it is the terms
proportional to $|y_t|^2$ that are mainly responsible for driving $m_{H_2}^2
< m_{H_1}^2$ in the first place. Lowering the third family soft sfermion
masses will reduce the difference $m_{H_1}^2 - m_{H_2}^2$ and thereby lower $%
m_{A^0}$. Fig.~\ref{fig4} shows the RG running of this difference from $M_{G}$ to $m_Z$
numerically for the three different sfermion soft mass matrices considered in
this paper, for $\tan\beta = 50$ and typical values $m_0 = m_{1/2} = 500$ GeV. Also shown is the value of $m_{A^0}$ in each case.

\FIGURE[tbp]{\includegraphics[width=\textwidth]{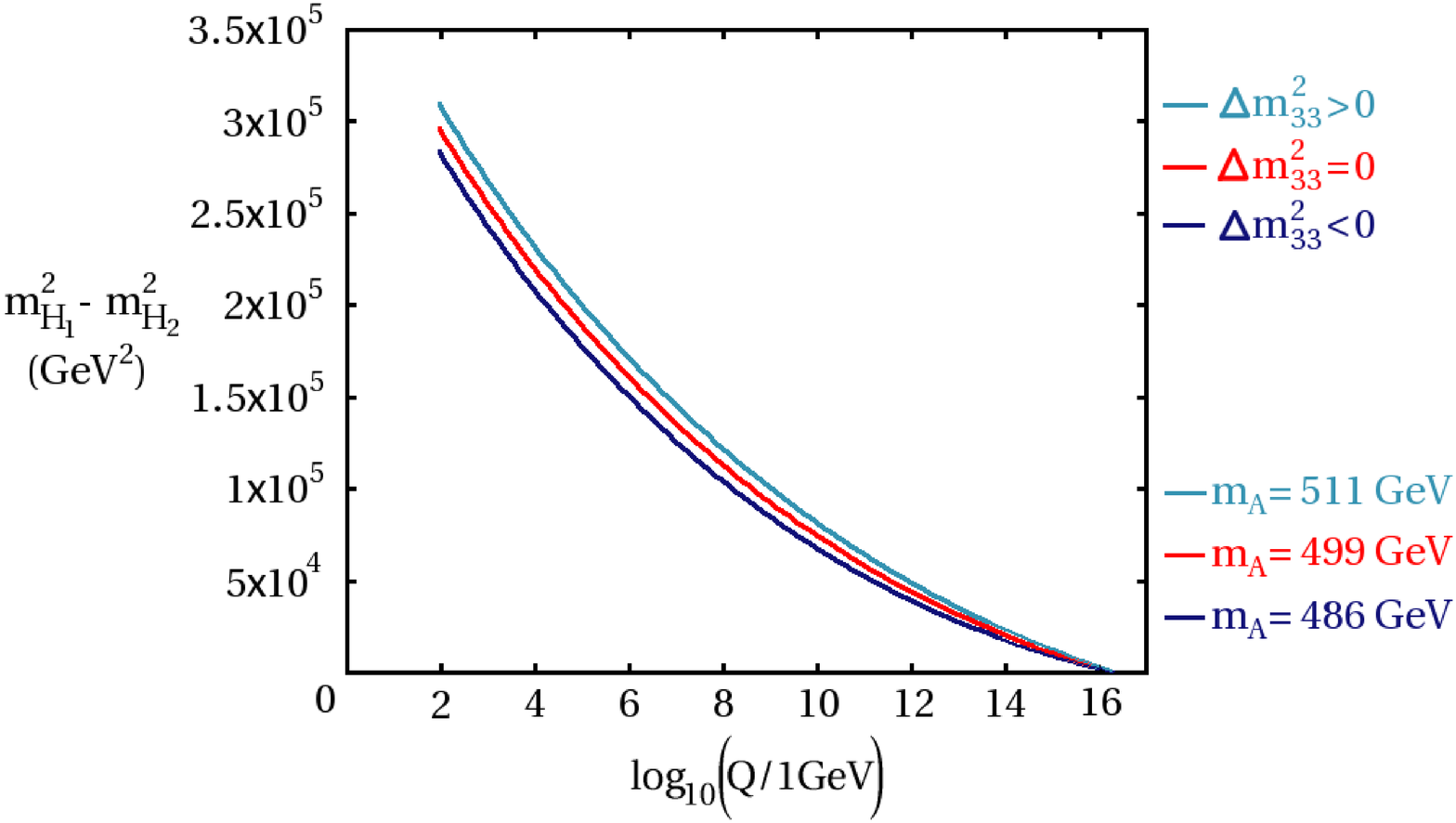}
\caption{\textit{This plot shows renormalisation group evolution of the
difference $m_{H_1}^2 - m_{H_2}^2$ from the GUT scale down to the EWSB scale
for the three different boundary conditions on the sfermion soft mass matrices with $m_0 =
m_{1/2} = 500$ GeV, $A_0 = 0$, $\tan\protect\beta = 50$ and $\protect\mu > 0$%
. The cyan(light) line corresponds to $\protect\delta m^{2}_{\tilde{f}_{33}}(m_G) =
m_0^2 \protect\delta m^2$, the red(medium) line to $\protect\delta m^{2}_{\tilde{f}%
_{33}}(m_G) = 0$, and the dark blue(dark) line to $\protect\delta m^{2}_{\tilde{f%
}_{33}}(m_G) = - m_0^2 \protect\delta m^2$}}
\label{fig4}}

The opposite effects to those described above pertain to Fig.~\ref{fig3}(c).
There is no boundary where $|\mu|^2 = 0$ below values of $m_0 \sim 8$ TeV or 
higher and the $A^0$ resonance is slightly shifted towards higher $m_{1/2}$. As a result, the neutralino relic abundance at resonance is increased due to the larger values of $\mu$ and $m_{\tilde{\chi}^0_1}$ relative to the case with no sfermion splitting.

\section{Summary and Conclusions\label{summary}}

Supersymmetric phenomenology is very sensitive to the soft SUSY breaking
parameters which determine the spectrum of the new supersymmetric states.
These parameters are strongly constrained by the need to avoid large flavour
changing neutral currents and this has led to the construction of several
distinct classes of model for supersymmetry breaking which are capable of
solving the SUSY family problem. In this paper we have explored some of the
implications of an alternative solution following from a non-Abelian family
symmetry. This is perhaps a more attractive solution than those based on
gravity mediation, gauge mediation or anomaly mediation because the solution
arises as a natural byproduct of a theory of fermion masses. In the case
there is an $SU(3)$ family symmetry the soft SUSY breaking masses of the
three family members in a given representation of the Standard Model are
degenerate up to $SU(3)$ breaking effects and this is sufficient to avoid
large FCNC. However this symmetry must be strongly broken to generate the
third family of fermion masses and this inevitably leads to a breaking of
the sfermion degeneracy, splitting the third family from the first two
families. Although this breaking is small enough to avoid unacceptably large
FCNC it does lead to significant changes in the phenomenology, particularly
in the radiative generation of electroweak breaking and in the dark matter
abundance following from the existence of a stable lightest supersymmetric
particle. We have explored these effects in detail for the case that an
underlying GUT guarantees the initial degeneracy of all the squarks and
sleptons of a given family.

The main conclusion is that even the the small splitting of the degeneracy
of the third family of the size indicated by the fermion mass structure
leads to significant changes from the CMSSM phenomenology that has been
widely used as a benchmark for future SUSY searches. In particular a
reduction of the third family squark masses leads to a reduction in the
radiative corrections that are needed to trigger electroweak breaking and
this in turn extends the region excluded by the WMAP constraints.
Associated with this is the fact that the region where the $\mu $ mass is
small, close to the electroweak exclusion region, moves in the $(m_{1/2},m_{0})$ plane. In this region the Higgsino component of the LSP is
enhanced and this significantly affects the LSP annihilation rate. This opens
up a new region of parameter space where the LSP residual abundance is able
to explain the dark matter abundance. It will be important to explore this
region too in future searches for supersymmetry.

The effects explored here are the minimal ones to be expected in the case
the family problem is solved by a non-Abelian family symmetry. Although we
have motivated the study in the context of a specific $SU(3)$ family
symmetry it is likely they have more general applicability as the magnitude
of the effects follow from the need to generate the large masses of the
third family of quarks and leptons and they are further constrained by the
need to suppress FCNC. As we have discussed above, a non-Abelian family
solution to the family problem allows for even more significant changes from
the CMSSM boundary conditions because, when the underlying GUT is broken,
there may be significant splitting between different Standard Model
representations provided each representation is separately nearly degenerate
in family space. These effects will be explored elsewhere~\cite{Ramage:2004wa}.

\acknowledgments
We would like to thank B. Allanach and J. Ellis for useful discussions. One of us (GGR) would like to thank S. King for helpful discussions. MR thanks the authors of Micromegas for all their help and is grateful to PPARC for studentship and travel support. This work was partly supported by the EU network, Physics Across the Present Energy Frontier HPRV-CT-2000-00148.

\bibliography{ramageross3}

\end{document}